\documentclass[aip, amsmath,amssymb, reprint,]{revtex4-1}

\usepackage{graphicx}
\usepackage{dcolumn}
\usepackage{bm}

\usepackage[utf8]{inputenc}
\usepackage[T1]{fontenc}
\usepackage{mathptmx}
\usepackage{etoolbox}
\usepackage{xcolor}

\makeatletter
\def\@email#1#2{
 \endgroup
 \patchcmd{\titleblock@produce}
  {\frontmatter@RRAPformat}
  {\frontmatter@RRAPformat{\produce@RRAP{*#1\href{mailto:#2}{#2}}}\frontmatter@RRAPformat}
  {}{}
}
\makeatother
\begin{document}

\title{Noise tailoring, noise annealing and external noise injection strategies in memristive Hopfield neural networks}

\author{J\'anos Gerg\H{o} Feh\'erv\'ari}
\affiliation{Department of Physics, Institute of Physics, Budapest University of Technology and Economics, Műegyetem rkp. 3., H-1111 Budapest, Hungary}
\affiliation{ELKH-BME Condensed Matter Research Group, M\H{u}egyetem rkp. 3., H-1111 Budapest,\unpenalty~Hungary}

\author{Zolt\'an Balogh}
\affiliation{Department of Physics, Institute of Physics, Budapest University of Technology and Economics, Műegyetem rkp. 3., H-1111 Budapest, Hungary}
\affiliation{ELKH-BME Condensed Matter Research Group, M\H{u}egyetem rkp. 3., H-1111 Budapest,\unpenalty~Hungary}

\author{T\'imea N\'ora T\"or\"ok}
\affiliation{Department of Physics, Institute of Physics, Budapest University of Technology and Economics, Műegyetem rkp. 3., H-1111 Budapest, Hungary}
\affiliation{Institute of Technical Physics and Materials Science, Centre for Energy Research, Konkoly-Thege M. \'{u}t 29-33, 1121 Budapest, Hungary}

\author{Andr\'as Halbritter}
\email{halbritter.andras@ttk.bme.hu}
\affiliation{Department of Physics, Institute of Physics, Budapest University of Technology and Economics, Műegyetem rkp. 3., H-1111 Budapest, Hungary}
\affiliation{ELKH-BME Condensed Matter Research Group, M\H{u}egyetem rkp. 3., H-1111 Budapest,\unpenalty~Hungary}

\begin{abstract}
The commercial introduction of a novel electronic device is often preceded by a lengthy material optimization phase devoted to the suppression of device noise as much as possible. The emergence of novel computing architectures, however, triggers a paradigm change in noise engineering, demonstrating that a non-suppressed, but properly tailored noise can be harvested as a computational resource in probabilistic computing schemes. Such strategy was recently realized on the hardware level in memristive Hopfield neural networks delivering fast and highly energy efficient optimization performance. Inspired by these achievements we perform a thorough analysis of simulated memristive Hopfield neural networks relying on realistic noise characteristics acquired on various memristive devices. These characteristics highlight the possibility of orders of magnitude variations in the noise level depending on the material choice as well as on the resistance state (and the corresponding active region volume) of the devices. Our simulations separate the effects of various device non-idealities on the operation of the Hopfield neural network by investigating the role of the programming accuracy, as well as the noise type and noise amplitude of the ON and OFF states. Relying on these results we propose optimized noise tailoring, noise annealing, and external noise injection strategies.
\end{abstract}

\maketitle

\section{Introduction}

Memristive crossbar arrays are promising candidates as the hardware components of artificial neural networks,\cite{Zidan2018,Xia2019,Mehonic2020,Sebastian2020,Ielmini2023} including advanced applications in feed-forward neural networks,\cite{Ambrogio2018,Li2018b,Wang2019a,Burr2019} convolutional layers,\cite{Li2018a,Wang2019b} 3D architectures,\cite{Seok2014,Wu2017,Li2017,Lin2020} unsupervised neural networks,\cite{Serb2016,WeiLu2017,Wang2018} as well as recurrent neural networks.\cite{Wang2019b,Li2019} In these applications the tunable conductance of a memristor unit encodes a synaptic weight in the network, and once the properly trained weights are programmed to each memristor cell, the memristive crossbar array is able to perform the vector-matrix multiplication, i.e. the key mathematical operation of the network inference in a single time-step.\cite{Xia2019,Zidan2018,Ambrogio2018} This equips the artificial neural networks with a highly energy efficient hardware component compared to software solutions, where the evaluation of the input vector at a layer with $N$ neurons requires $N^2$ multiplication operations. In most of the neural network applications the highest resolution of the memristive synaptic weights is desirable,\cite{Rao2023} and therefore the memristor non-idealities, like their programming inaccuracy or their stochastic noise properties should be eliminated as much as possible. A special class of the memristive networks, however, relies on probabilistic optimization,\cite{Kim2023,Li2023,Fahimi2021,cai2020} where it is well known that tunable stochasticity, such as customizable device noise, is not a disturbing factor, but a useful computational resource. Similar strategy was recently experimentally realized in $60\times 60$ memristive Hopfield neural networks (HNNs),\cite{cai2020} demonstrating the efficient solution of max-cut graph segmentation problems, and delivering over four orders of magnitude higher solution throughput per power consumption than digital or quantum annealing approaches. 

Inspired by these achievements, we perform a thorough analysis of simulated memristive Hopfield neural networks putting a key emphasis on the effect of the device noise on the network operation. To this end, first the realistic noise characteristics of memristive devices\cite{santa2021,Balogh2021,Santa2019,Posa2021} are discussed (Sec.~\ref{noise}), and a general noise model, describing the conductance-dependent noise characteristics in the filamentary and broken filamentary regimes is proposed. 
Afterwards, various benchmark max-cut problems are solved by simulated memristive HNNs (Sec.~\ref{simulation}), relying on the proposed noise model. These simulations demonstrate rather well-defined relative noise values, at which the network operation is optimized, regardless of the network size and the type of the noise spectrum. We also demonstrate a simplified, easily implementable double-step noise annealing scheme (Sec.~\ref{annealing}), which further enhances the convergence probability of the network. These optimized noise levels, however, are at the top border of the experimentally observed noise amplitudes, which raises the need for external injection of stochasticity (Sec.~\ref{external}). For the latter, two strategies are tested, including external noise injection and the introduction of chaotic behavior through the self-feedback of the neurons in the network.\cite{Huang2020} Finally, the effect of further device-non-idealities are tested separately (Sec.~\ref{furthernonidealities}), analyzing the effect of the programming inaccuracy, and the finite OFF-state conductance. The presentation of all these results is preceded by the brief overview of Hopfield neural networks, and their implementation by memristive crossbar arrays (Sec.~\ref{HNN}). 

\section{Memristive Hopfield Neural Networks}
\label{HNN}

\subsection{Hopfield Networks}

\begin{figure}
    \centering
    \includegraphics[width=0.88\columnwidth]{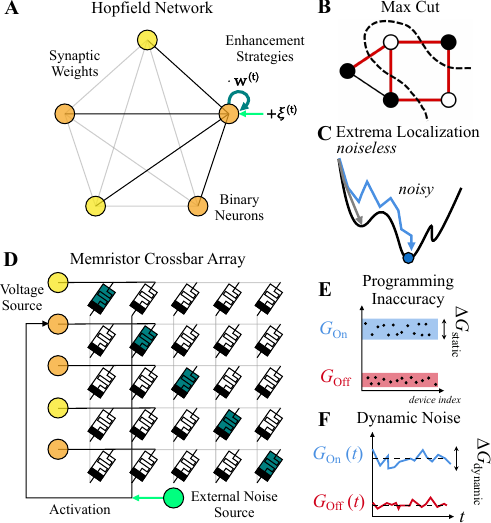}
    \caption{(A) Illustration of a Hopfield neural network with five neurons. The orange (yellow) circles illustrate the ‘+1’ (‘-1’) binary states, whereas the lines represent the synaptic weights between the neurons. A HNN excludes self-connections, however, a self-connection with negative weight (dark green arrow) introduces chaotic nature to the network operation, which helps to find the global solution. Stochasticity can be also introduced by the temporal variation (noise) of the synaptic weights, as well as by external noise injection (light green arrow). (B) Illustration of the max-cut problem: the goal is to find a partitioning of the vertices into two adjacent sets so that the total number of crossing edges between the two sets (red lines with dashed-line cut) is maximal. (C) Illustration of the energy landscape of a HNN. A noiseless operation with the conventional update rule may yield dead-ends in local minima (grey line). Properly tailored stochasticity, however, helps escaping from the local minima, and finding the global solution (blue line). (D) Experimental realization of the discrete HNN by a memristor crossbar array. The $V_i=\pm|V|$ voltage inputs at the horizontal lines represent the states of the neurons, which are updated according to the $I_j$ current outputs at the vertical lines, the latter representing the $a_j^{(t)}$ activation. The synaptic weights are encoded in the $G_{i,j}$ conductance matrix of the memristors in the crossbar. In a conventional HNN the lack of self-connections is represented by the $\approx 0$ conductance values at the diagonal of the crossbar (dark green memristors). The light green arrow illustrates the possibility of external noise injection. (E) In a memristive HNN the '1' and '0' synaptic weights are encoded in $G_\mathrm{ON}$ and $G_\mathrm{OFF}$ conductance values. These, however, exhibit device-to-device variations described by the $\Delta G_\mathrm{static}$ variance. (F) The stochastic temporal variation (i.e. the noise) of the $G_\mathrm{ON}$ and $G_\mathrm{OFF}$ conductance values also introduces a device non-ideality described by the $\Delta G_\mathrm{dynamic}$ variance. The proper tailoring of the noise, however, aids the network operation.}
    \label{fig:0}
\end{figure}

The Hopfield Neural Network (HNN) introduced by John Hopfield,\cite{hopfield1982} was shown to be capable of solving complex problems by Hopfield and Tank\cite{hopfield1986} and has been used for optimization ever since.\cite{smith1999} A Hopfield Network's main allure is its simplicity and immense power to provide reasonable solutions for high complexity problems. The network consists of fully connected binary neurons (without self-connections), the $\underline{\underline{W}}$ synaptic weight matrix encodes the optimization problem and the $\underline{x}$ state of the neurons represent the possible states of the system including the desired solution(s) (see Fig.~\ref{fig:0}A). The network is operated in an iterative fashion: at each time step $t$ the activation
\begin{equation}
      \underline{a}^{(t)} = \underline{\underline{W}} \cdot \underline{x}^{(t)}
\label{update}
\end{equation} is calculated. Then an index $j$ is picked at random and a single neuron is updated according to the rule
\begin{equation}
       x_j^{(t+1)} =
\begin{cases}
+1 \quad \mbox{if} \ a_j^{(t)} \geq \theta_j \text{,} \\
-1 \quad \mbox{if} \ a_j^{(t)} < \theta_j \text{,}
\end{cases}
\label{activation}
\end{equation} where $\theta_j$ is a component of a predefined threshold vector $\underline{\theta}$.
It can be shown, that this simple update rule decreases the energy function 
\begin{equation}E^{(t)} \left(\underline{x}^{(t)}; \underline{\underline{W}}, \underline{\theta} \right) = - \frac{1}{2} \left(\underline{x}^{(t)}\right)^\mathrm{T} \underline{\underline{W}} \ \underline{x}^{(t)} + \underline{\theta} \ \underline{x}^{(t)}
\label{energyfunction}
\end{equation} in every iteration step ($E^{(t+1)}\le E^{(t)}$). Due to this property, Hopfield neural networks are widely used to solve complex problems, which can be encoded in the form of the effective energy function in Eq.~\ref{energyfunction}. This can be applied in an associative memory scheme,\cite{hopfield1982} where $\underline{\underline{W}}$ and $\underline{\theta}$ are set such, that each local minimum of $E(\underline{x})$ encodes a predefined pattern (e.g. images), and the update rule drives the system from an arbitrary initial state $\underline{x}^{(0)}$ to the closest local minimum, i.e. the network finds the predefined pattern most similar to the initial state. Alternatively, the Hopfield neural network may find the global solution of a complex problem, like the max-cut graph segmentation problem. 

\subsection{Max-cut problem}
The NP-hard max-cut problem is formulated for an arbitrary undirected $G(V,E)$ graph with $V$ vertices and $E$ edges. The goal is to find a partitioning of $V$ into two adjacent sets $S$ and $K$ so that the total weight of crossing edges between the two sets is maximal (see Fig.~\ref{fig:0}B).\cite{wu2001} Though abstract at first sight, many practical problems can be mapped to the max-cut, such as the conflict-graph formulation of the layer assignment problem in very large scale integration (VLSI) design, where the position of functional blocks is optimized in a multilayer chip.\cite{frauke2011} If a graph is given by its adjacency matrix $\underline{\underline{A}}$, a cut can be encoded by $\underline{x}$, simply as
\begin{equation}
       x_k = \begin{cases}
        +1 \quad \mbox{if} \ V_k \in S \text{,} \\
        -1 \quad \mbox{if} \ V_k \in K \text{.}
    \end{cases}
    \label{adjacency}
\end{equation}
and the maximum cut can be found by minimizing the $E(\underline{x}; -\underline{\underline{A}}, \underline{0})$ energy function (Eq.~\ref{energyfunction}). For an unweighted graph, the absolute value of the energy is proportional to the number of edges running between $S$ and $K$, so the problem is directly addressable by a HNN.\cite{wu2001} In that case, however, the global minimum of the energy function is to be found, while the conventional operation of a Hopfield neural network would yield dead ends in each local minima of the energy landscape. This problem can be eliminated by introducing proper stochasticity to the network, like a finite noise, which helps escaping from the local minima, and which is reduced as the states evolve towards the global solution (see the illustration in Fig.~\ref{fig:0}C).  

\subsection{Hardware implementation of HNNs by memristive crossbar arrays}

The so-called crossbar structure is a popular scheme for physical matrix realization via memristors.\cite{Xia2019,li2021} As seen in Fig. \ref{fig:0}.D it is essentially a set of horizontal and vertical wires (word and bit lines) with a memristor placed at each meeting point of the lines. Operating this arrangement in the linear, subthreshold regime of the memristive units, the output current vector at the bit lines is obtained as the product of the input voltage vector at the word lines and the conductance matrix of the memristors at the crosspoints, $I_j = \sum_{i} G_{i,j} \cdot V_i$. Once the proper conductance weights are programmed to the crossbar, the vector-matrix multiplication is performed on the hardware level within a single clock-cycle. This scheme is also implementable for Hopfield neural networks, where the diagonal values of the $G_{i,j}$ conductance matrix are zero due to the lack of self-connections in the HNN. 

The special case of the max-cut problem is mathematically formulated by a weight matrix which is '1' or '0' if the proper vertices are connected or non-connected. This problem can be mapped to a memristive HNN by setting a constant $G_\mathrm{ON}$ conductance and $G_\mathrm{OFF}\ll G_\mathrm{ON}$ conductance instead of the $1$ and $0$ values, respectively. The $x_i$ binary state vector elements are represented by $V_i^{(t)}=x_i^{(t)}\cdot |V|$ input voltages at the crossbar word lines. This scheme was experimentally realized in Ref.~\citenum{cai2020} using memristive HNNs up to $60\times60$ matrix sizes.

\subsection{Non-idealities and stochasticity in memristive HNNs}

Figures~\ref{fig:0}E,F demonstrate the key device non-idealities in a memristive Hopfield neural network: the temporal stochastic variation (noise) of the programmed $G_\mathrm{ON}$ and $G_\mathrm{OFF}$ conductances (F), as well as the programming inaccuracy, i.e. the device-to-device variation of the time-averaged $\overline{G_{i,j}(t)}$ conductances for the memristor cells programmed to the same ON/OFF binary state. These non-idealities are measured by the temporal variance ($\Delta G_\mathrm{dynamic}$) and the device-to-device variance ($\Delta G_\mathrm{static}$) around the ideal 
$G_\mathrm{ON}$ and $G_\mathrm{OFF}$ values. Furthermore it is noted, that $G_\mathrm{ON}$ can be chosen arbitrary in the mapping of the '0' and '1' synaptic weights to the $G_\mathrm{ON}$ and $G_\mathrm{OFF}$ conductance, however, a finite $G_\mathrm{OFF}$ already represents a device non-ideality, which may modify the operation of the network. Furthermore, finite wire resistances or nonlinear $I(V)$ characteristics may also be considered as a non-idealities,\cite{cai2020} however, these two non-idealities are not considered in our following analysis. 

Among these non-idealities the noise plays a distinguished role, as it not necessarily hampers the network operation, but a properly tailored noise may help to find the global solution. However, the injection of external stochasticity might also become necessary once the internal noise of the memristor elements in the crossbar array is not enough large for the optimal network operation. The latter possibility is illustrated by the light green arrows in Figs.~\ref{fig:0}A,D as well as by the dark green arrow in Fig.~\ref{fig:0}A respectively illustrating external noise injection and the introduction of chaotic behavior via a negative self-feedback of the neurons.\cite{cai2020,Huang2020}   

\section{Realistic noise properties of memristive devices}
\label{noise}

In the following we analyze the realistic noise characteristics of various memristive systems (Fig.~\ref{fig:1}), which are considered as a key ingredient of the memristive HNNs' operation. 

\subsection{Typical noise spectra of memristive units}

\begin{figure}
    \centering
    \includegraphics[width=0.9\columnwidth]{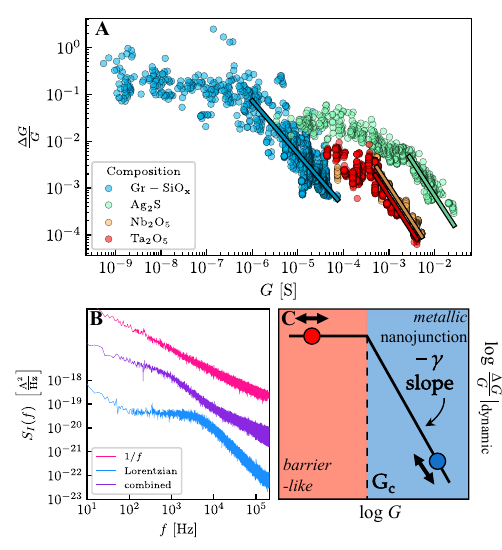}
    \caption{(A) Relative conductance noise as a function of device conductance for a variety of memristive materials. The data for Ag$_2$S (green), Ta$_2$O$_5$ (red) and Nb$_2$O$_5$ (orange) memristive devices are reproduced from Refs.~\citenum{Santa2019,santa2021,Balogh2021}, recalculating the integrated noise amplitudes for the $[f_A,f_B]=[10\,\mathrm{Hz},250\,\mathrm{kHz}]$ band. For these material systems the validity of the diffusive noise model with volume-distributed fluctuators was verified in the high conductance regime, the lines with the corresponding colors represent the best fitting trends with the corresponding $\gamma=3/2$ exponent.\cite{Santa2019,santa2021,Balogh2021} At somewhat smaller conductances a rather narrow ballistic conductance region is observed with significantly shallower, $\gamma=1/4$ exponent.\cite{Santa2019,santa2021,Balogh2021} Finally, in the sub-conductance-quantum interval a broken filamentary regime is observed with $\approx \mathrm{constant}$ relative noise level, which is best resolved for the Ag$_2$S system.\cite{Balogh2021} The blue data represent new measurements on graphene-SiO$_x$-graphene lateral devices, using the sample preparation protocol as in Ref.~\citenum{Posa2017,Posa2021}. Here the switching relies on the voltage-controlled transitions between well-conducting crystalline and poorly conducting amorphous regions.\cite{Torok2022,JTour2012} The low conductance $\approx \mathrm{constant}$ and high conductance $\sim G^{-\gamma}, \gamma=1.13$ dependencies are clearly seen for this system spanning 5 orders of magnitude (3 orders of magnitude) range along the conductance (relative noise) axis. The $G_C$ crossover conductance is well below $G_0$ indicating a barrier-like component even in the metallic regime. (B) Illustrative Lorentzian (blue) 1/f-type (pink) and mixed (purple) noise spectra measured on Ta$_2$O$_5$ devices. The bottom curve is on true scale, while the middle and top curves are artificially shifted upwards by one and two orders of magnitude. (C) Proposed noise model with constant relative noise in the barrier-like regime ($G<G_C$) and $\sim G^{-\gamma}$ relative noise in the metallic nanojunction regime ($G>G_C$), the $G_\mathrm{ON}$ (red circle) and $G_\mathrm{OFF}$ (blue circle) conductances can be set to arbitrary positions along the noise model.}
    \label{fig:1}
\end{figure}

The $S_I(f)$ spectral density of the current noise is defined as the $(\Delta I)^2|_{f_0,\Delta f}$ mean squared deviation of the current within a narrow $\Delta f$ band around the central frequency $f_0$ normalized to the bandwidth, $S_I(f_0)=(\Delta I)^2|_{f_0,\Delta f}/\Delta f$, but practically $S_I(f)$ is calculated from the absolute value squared of the Fourier transform of the measured $I(t)$ fluctuating current signal.\cite{Balogh2021} In memristive devices $S_I(f)$ typically exhibits a Lorentzian spectrum (blue curve in Fig.~\ref{fig:1}B), or a 1/f-type spectrum (pink curve in Fig.~\ref{fig:1}B), or the mixture of these two (purple curve in Fig.~\ref{fig:1}B). In the former case the noise is dominated by a single fluctuator introducing a steady-state resistance fluctuation with a typical time constant $\tau_0$, yielding a spectrum which is constant at $2\pi f<\tau_0^{-1}$, and decays with $1/f^2$ at $2\pi f>\tau_0^{-1}$.\cite{Balogh2021} If multiple fluctuators with different time constants contribute to the device noise, the Lorentzian spectra of the individual fluctuators sums up to a spectrum with $S_I\sim f^{-\beta}$, where $\beta$ is usually close to unity (pink noise, pink curve in Fig.~\ref{fig:1}B).\cite{Balogh2021} Alternatively, a single fluctuator positioned at the device bottleneck may dominate the device noise, but a larger ensemble of more remote fluctuators may also give a significant contribution.\cite{Balogh2021} This situation yields the mixture of Lorentzian and 1/f-type noise (purple curve in Fig.~\ref{fig:1}B). Without any steady-state resistance fluctuations, still, a finite thermal noise is observed, the latter exhibiting a constant (frequency-independent) spectrum (white noise). Integrating the current noise for the frequency band of the measurement, the mean squared deviation of the current is obtained in this band, $(\Delta I)^2=\int_{f_A}^{f_B}S_I(f)\mathrm{d}f$.

\subsection{Proper metrics of the noise characteristics}

At low enough sub-threshold voltages the memristive conductances exhibit steady-state fluctuations, i.e. the applied voltage is only used for the readout of the noise, but it does not excite any fluctuations. In this case $(\Delta I)^2=(\Delta G)^2\cdot V^2$ holds according to Ohm's law, i.e. the voltage-dependent current fluctuation is not a good measure of the noise properties. The $\Delta I/I$ relative current fluctuation, however, is already a voltage-independent metric of the fluctuations, which equals the relative fluctuation of the conductance or the resistance in the linear regime, $\Delta I/I=\Delta G/G=\Delta R/R$.\cite{Balogh2021} This metric will be used throughout the paper to describe the noise characteristics, where $(\Delta G/G)_\mathrm{dynamic}$ describes the relative temporal fluctuations of a certain element of the memristor conductance matrix $G_{i,j}(t)$. It is noted, that $(\Delta G/G)_\mathrm{dynamic}$ depends on the bandwidth. The high-frequency cutoff is determined by the integration time of the current readout ($\tau_\mathrm{readout}=2\,\mu$s in our simulation yielding $f_B=1/2\tau_\mathrm{readout}=\,250$kHz), whereas the $f_A=10\,$Hz bottom end of the frequency band is determined by the time-period for which the network is operated ($0.1\,$s in our simulation corresponding to $10000$ iteration steps, and a $4\tau_\mathrm{readout}$ waiting time between the current readout events, simulating the finite time of the neural updates and the multiplexing). Increasing the number of iterations steps would naturally increase the $(\Delta G/G)_\mathrm{dynamic}$, but this dependence is characteristic to the nature of the noise spectrum. In case of a Lorentzian spectrum the noise amplitude does not really depend on the bandwidth, once the $(2\pi \tau_0)^{-1}$ characteristic frequency of the fluctuator is well inside the band. This is consistent with the $(\Delta G)^2\sim \arctan (2\pi f\tau_0)|_{f_A}^{f_B}$ relation for the Lorentzian spectrum. The other experimentally relevant, $1/f$-type spectrum yields $(\Delta G)^2\sim\ln (f_B/f_A)$, which is also a very weak dependence on the bandwidth, yielding only a $\approx30\%$ increase of $(\Delta G/G)_\mathrm{dynamic}$ once the number of iteration steps is increased from $10^4$ to $10^7$, i.e. the above bandwidth is increased by three orders of magnitude. According to these considerations, the results of our following simulations are rather weakly dependent on our specific choice for the bandwidth.

\subsection{Variation of the noise with the device conductance}

Several studies have pointed out, that the relative noise amplitude of a memristive device exhibits a strong and specific dependence on the device conductance, i.e. the multilevel programmability is accompanied by the tuning of the relative noise level.\cite{Ielmini2010,Soni2010,Fang2013,Ambrogio2014,Ambrogio2015,Yi2016,Puglisi2018,piros2020,Santa2019,santa2021,Lee2023} Fig.~\ref{fig:1}A shows four examples for this behavior, demonstrating the conductance-dependent noise characteristics of Ag$_2$S (green),\cite{Santa2019,Balogh2021} Ta$_2$O$_5$ (red),\cite{santa2021} Nb$_2$O$_5$ (orange)\cite{santa2021} and SiO$_x$ (blue) memristive units integrated for the same $10\,$Hz-$250\,$kHz frequency band. It is clear, that the overall noise amplitude, the characteristic conductance range of the operation, as well as the dependence of $\Delta G/G$ on the conductance is a kind of a device fingerprint, exhibiting significant differences between various material systems. However, a rather general trend of the noise characteristics can be identified: in the low-conductance region of the operation regime $\Delta G/G$ is very weakly dependent on the conductance, whereas in the high conductance region a strong $\Delta G/G\sim G^{-\gamma}$ power-law dependence is typical. 

In the latter case, a metallic filamentary conduction is envisioned, where the relative noise amplitude obviously increases as the filament diameter is reduced.\cite{Balogh2021} A rather generally observed tendency is related to the volume-distributed fluctuators in a diffusive filament, where $\Delta G/G\sim G^{-3/2}$ was obtained from theoretical considerations.\cite{Santa2019,santa2021} This is also confirmed by the experimental data in Fig.~\ref{fig:1}A, where the validity of the $\gamma=3/2$ exponent was approved for the Ag$_2$S, Ta$_2$O$_5$ and Nb$_2$O$_5$ systems, \cite{Santa2019,santa2021} whereas our new data on SiO$_x$ memristors exhibits a somewhat shallower dependence with $\gamma=1.13$ (see the dashed lines representing the best fitting tendecies with the given $\gamma$ exponents). It is noted, however, that the $\gamma$ exponent may depend on the transport mechanism, the device geometry, the dimensionality (2D/3D devices), as well as the distribution of the fluctuators (single or multiple fluctuators, surface or volume distributed fluctuators, etc.).\cite{Balogh2021} 

In contrast, the saturated noise characteristics in the low conductance regime are attributed to broken filaments, where a barrier-like transport is envisioned. In the simplest case of a tunnel barrier the $G=A\cdot\exp(-\alpha\cdot d)$ relations yields a conductance-independent $\Delta G/G=\alpha \cdot \Delta d$ relative conductance noise for a constant $\Delta d$ fluctuation of the barrier width.\cite{Balogh2021} More complex transport phenomena, like the Frenkel-Poole mechanism,\cite{Slesazeck2015} or a hopping-type transport\cite{Gao2015} require more sophisticated descriptions, but the overall trend, i.e. the independence, or the very weak dependence of $\Delta G/G$ on $G$ is left unchanged due to the exponential dependence of the conductance on a relevant fluctuating parameter. 

According to these considerations, in the following simulations we rely on a simplified noise model (see Fig.~\ref{fig:1}C), where $\Delta G/G$ is constant below a certain threshold conductance $G_C$ (see the red barrier-like regime in Fig.~\ref{fig:1}C), whereas a general $\Delta G/G\sim G^{-\gamma}$ power-law dependence is considered at $G>G_C$ (see the blue metallic nanojunction regime in Fig.~\ref{fig:1}C). The $G_\mathrm{OFF}$ and 
$G_\mathrm{ON}$ conductances of the memristive HNN can be fixed at arbitrary positions along this noise model, as demonstrated by the red and blue circles in Fig.~\ref{fig:1}C. This simplified model has three free parameters, the $G_C$ threshold, the $\gamma$ slope, and the $\Delta G/G$ relative fluctuation in the barrier-like regime. Note, that according to the experimental results in Fig.~\ref{fig:1}A the latter can reach a few tens of percents, $G_C$ is not necessarily, but reasonably close to the $G_0=2e^2/h$ conductance quantum unit, whereas the variation of $\Delta G/G$ can span up to three orders of magnitude in the metallic nanojunction regime. For the exponents $\gamma=1.13-1.5$ values are observed, however, we emphasize that  fundamentally different slopes are also possible.\cite{Balogh2021} 

\section{Simulation of Memristive HNNs with Realistic Noise Characteristics}
\label{simulation}

We have simulated memristive HNNs with realistic noise characteristics relying on the standardized Biq Mac Library\cite{wiegele2007} which provides exact globally optimal energies for Max-Cut instances of undirected and unweighted graphs in the sizes $n \in [60,80,100]$. Following the results of Ref.~\citenum{cai2020}, we were studying Erdős-Rényi graphs with $50\%$ connection probability between the vertices. We have simulated the HNNs starting from $K=200$ randomly picked initial state vectors, and performing $N=10000$ iterations for each epoch.  The neurons were iterated in a predetermined random order. The $K$ runs are evaluated according to two figures of merit, the proportion of runs, where the network was in the globally optimal state at the $N^\text{th}$ step ($\mathbb{P}_\mathrm{conv}$ convergence probability), and the number of edges between the two subsets (i.e. the $C$ number of cuts) after $N$ iteration steps averaged for the K random initial vectors, $\overline{C}= \frac{1}{K} \sum_{i=1}^{K} C \left(\underline{x}_i^{(N)}\right)$.

Instead of the ideal '$1$' and '$0$' values of the $W_{i,j}$ weight matrix, realistic $G_{i,j}$ conductances of the memristive HNN were used in the simulations. 
At the matrix positions with a value of '$1$'in the original problem an average conductance of $G_\mathrm{ON}$ was applied, considering both $(\Delta G/G)_\mathrm{static}$ device-to-device variations and $(\Delta G/G)_\mathrm{dynamic}$ temporal fluctuations around this mean value. To simulate the latter, independent $G(t)$ time traces were generated for all the memristive elements in the ON state, using either Lorentzian, pink or white noise spectrum. Carson's theorem and method was applied\cite{carrettoni2010} to generate the $G(t)$ temporal noise traces (i.e. temporal conductance variations) from the chosen $S_G(f)$ noise spectrum. 

According to the HNN scheme, the diagonal elements of the crossbar matrix (self-connections) are set to exactly zero. This can be physically implemented by omitting the memristors at the diagonal positions either by switching off their transistor in a 1T1R arrangement,\cite{Cai2019,Strachan2018,Jiang2018} or by omitting their electroforming procedure. In Sec.~\ref{external}, however, we  also discuss the case of finite self-feedback, which introduces a chaotic nature to the network.

The offdiagonal elements of the weight matrix with values of '$0$' are generally represented by the $G_\mathrm{OFF}$ conductance, with the corresponding relative conductance noise. In a part of the simulations, however, $G_\mathrm{OFF}=0$ is applied according to the following considerations.

\subsection{Contribution of the OFF and ON state elements to the current output and the noise of the memristive HNN}
\label{noisecontributions}

In the following we provide simple considerations on the relative current and current fluctuation contributions of the ON and OFF state elements, which helps to identify the most relevant contributions.

(i) \textit{Relative current contribution of OFF-state memristive elements in the crossbar.} For bit line $j$ the number of '$1$' values in the original weight matrix is denoted by $d_j$ yielding an average value of $\overline{d_j}=(n-1)\cdot p_\mathrm{c}$ according to the random  connection probability between the $n$ vertices, for which $p_\mathrm{c}=0.5$ is applied in the following. 
The current contribution of the ON and OFF state elements in a certain bit line $j$, however, also depends on the distribution of the '$+1$' and '$-1$' values in the $x_i$ state vector, which varies along the operation. For the ensemble average of the adjacency matrices with the same random connection probability, however, an ensemble-averaged current can be calculated
\begin{equation}
\overline{I_j}=\underbrace{\sum_i x_i \cdot |V|\cdot p_\mathrm{c}\cdot  G_\mathrm{ON}}_{\overline{I_\mathrm{ON,j}}}+\underbrace{\sum_i x_i \cdot |V| \cdot (1-p_\mathrm{c})\cdot  G_\mathrm{OFF}}_{\overline{I_\mathrm{OFF,j}}},
\label{IonIoff}    
\end{equation}
from which the $\overline{I_\mathrm{OFF,j}}/\overline{I_\mathrm{ON,j}}=(G_\mathrm{OFF}/G_\mathrm{ON})*(1-p_\mathrm{c})/p_\mathrm{c}$
ratio gives an indication on the OFF and ON state memristors' relative current contribution in column $j$. For the special case of $p_\mathrm{c}=0.5$ this simplifies to $\overline{I_\mathrm{OFF}}/\overline{I_\mathrm{ON}}=G_\mathrm{OFF}/G_\mathrm{ON}$. This demonstrates, that at large enough ON/OFF conductance ratio (e.g. $G_\mathrm{ON}/G_\mathrm{OFF}>100$), the replacement of $G_\mathrm{OFF}$ by zero is a reasonable simplification for a densely connected graph. Later on (Sec.~\ref{finiteoff}), we numerically analyze how a nonzero $G_\mathrm{OFF}$ value modifies the network operation at moderate ON/OFF conductance ratios.  

(ii) \textit{Relative noise contribution of OFF-state memristive elements in the crossbar.} Whereas the current in a certain bit line strongly depends on the actual $x_i$ state vector values, the mean squared deviation of the current is independent of that, and can be exactly deduced, once the $d_j$ number of ON-state elements in column $j$ is known:
\begin{align}
\label{reloffnoise1}
(\Delta I)^2_\mathrm{j}&=\sum_{i\ (i\ne j)}{(\Delta G)_\mathrm{j,i}^2\cdot |V_i|^2}=\\ \nonumber
&=\underbrace{(\Delta G)_\mathrm{OFF}^2\cdot(n-d_j-1)\cdot |V|^2}_{(\Delta I)^2_\mathrm{OFF,j}} +
\underbrace{(\Delta G)_\mathrm{ON}^2\cdot d_j\cdot |V|^2}_{(\Delta I)^2_\mathrm{ON,j}}.
\end{align}
From this the relative noise contributions of the OFF and ON state elements in bit line $j$ can be calculated considering our simplified noise model (Fig.~\ref{fig:1}C). First, we treat the \emph{mixed barrier-like and metallic} regime, where $G_\mathrm{ON}>G_\mathrm{C}>G_\mathrm{OFF}$, yielding: 
\begin{equation}\frac{\Delta I_\mathrm{OFF,j}}{\Delta I_\mathrm{ON,j}}=\frac{G_\mathrm{OFF}}{G_\mathrm{C}}\left(\frac{G_\mathrm{C}}{G_\mathrm{ON}}\right)^{1-\gamma}\cdot {\sqrt\frac{n-d_j-1}{d_j}}.
\label{reloffnoise2}
\end{equation}
Note, that the square-root term gives unity once $d_j$ is replaced by its average value at $50\%$ connection probability. This formula yields negligible OFF-state noise contribution for arbitrary $\gamma$, once $G_\mathrm{OFF}$ is chosen deep in the barrier-like regime ($G_\mathrm{OFF}/G_\mathrm{C}\ll 1$), whereas $G_\mathrm{ON}$ remains reasonably close to $G_\mathrm{C}$.

It is worth discussing another limit as well, where the entire crossbar is operated in the metallic nanojunction regime (see Fig.~\ref{fig:1}C), i.e. $G_\mathrm{ON}>G_\mathrm{OFF}>G_\mathrm{C}$ (\emph{pure metallic regime}). In the metallic nanojunction regime the memristive elements exhibit much more linear subthreshold $I(V)$ characteristics than in the barrier-like regime, which is a favorable property for the high-precision vector-matrix multiplication operation of the memristive crossbar. This limit yields 
\begin{equation}\frac{\Delta I_\mathrm{OFF,j}}{\Delta I_\mathrm{ON,j}}=\left(\frac{G_\mathrm{OFF}}{G_\mathrm{ON}}\right)^{1-\gamma}\cdot {\sqrt\frac{n-d_j-1}{d_j}},\label{reloffnoise3}\end{equation}
emphasizing the dominance of the OFF-state elements' noise contribution at any $\gamma>1$ value, i.e. for all the memristive units demonstrated in Fig.~\ref{fig:1}A. Furthermore, in this pure metallic operation regime the $G_\mathrm{ON}/G_\mathrm{OFF}$ conductance ratio is restricted to rather limited values spanning one order of magnitude in the diffusive regime of Ag$_2$S, Ta$_2$O$_5$ and Nb$_2$O$_5$ memristors, and less than two orders of magnitude in SiO$_x$ memristors (see Fig.~\ref{fig:1}A), which may distort the network operation compared to networks operated with orders of magnitude larger $G_\mathrm{ON}/G_\mathrm{OFF}$ ratios in the mixed barrier-like and metallic regimes (Eq.~\ref{reloffnoise2}), i.e. the choice of the operation regime is the tradeoff between high precision linearity and the proper representation of the '0' values in the weight matrix. In the following subsections, we discuss the results of our simulations for both operation regimes using max-cut benchmarks with $50\%$ connection probability and noise model with $\gamma=3/2$. 
Note, however, that the above formulae are also proper to discuss more general situations, including 
arbitrary conductances and $\gamma$ scaling exponents. Furthermore, arbitrary graphs (i.e. any $d_j$ and $n$ values) can also be analyzed, where the replacement of the applied dense graph with a sparse graph ($d_j/n\ll 1/2$) would yield the further enhancement of the OFF noise contribution. 

\subsection{Optimal noise level, and the role of the noise color}
\label{optimalnoise}

\begin{figure}
    \includegraphics[width= \columnwidth]{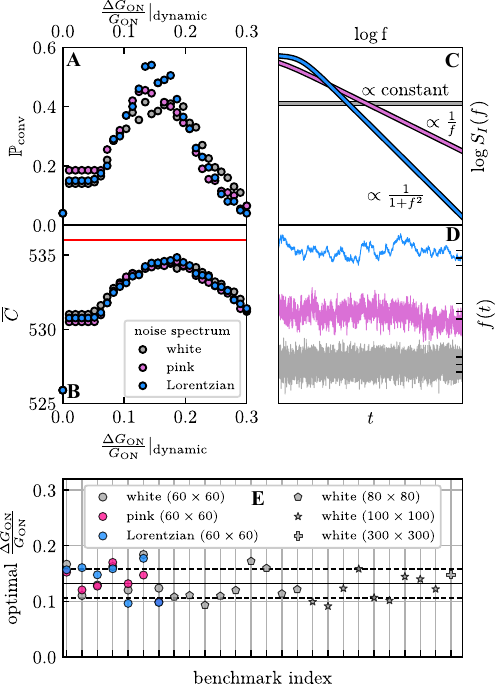}
    \caption{Simulation results for test problem g05\_60.0\,\cite{wiegele2007} with different dynamic noise spectra at various constant noise levels. (A) Convergence probability as a function of dynamic noise level for the three noise types. White noise, pink noise, and Lorentzian noise are respectively marked by grey, pink, and blue symbols in all panels. (B) $\overline{C}$ as a function of dynamic noise level for the three noise types. (C,D) Noise spectra and  example $G(t)$ traces generated from these spectra for the three noise types. 
    (E) Optimal noise level for various max-cut problems using graphs with randomly generated $50\%$ connection probabilities, and sizes of $60\times60$ (circles), $80\times80$ (pentagons), $100\times100$ (stars) and $300\times300$ (plus symbol).}
    \label{fig:2}
\end{figure}

We have simulated max-cut problems using graphs with $50\%$ connection probability and different sizes ($n=60,80,100,300$). First, the mixed barrier-like and metallic operation regime is analyzed (Eq.~\ref{reloffnoise2}) with $G_\mathrm{OFF}/G_\mathrm{ON}\ll 1$, and according to the above considerations $G_\mathrm{OFF}=0$ was chosen, whereas a finite $G_\mathrm{ON}$ value with variable noise was applied. The time-averaged conductance was the same for all the ON elements, i.e. $(\Delta G/G)_\mathrm{static}=0$ was applied. Simulations were run with three different noise types: Lorentzian noise (blue symbols in Fig.~\ref{fig:2}), $1/f$ noise (pink symbols), and white noise (grey symbols). The related noise spectra and time traces generated from these spectra are shown in Figs.~\ref{fig:2}C,D. For all three spectra, the $(\Delta G/G)_\mathrm{dynamic}$ metric was used to measure the relative noise.

Figs.~\ref{fig:2}A,B demonstrate the $\mathbb{P}_\mathrm{conv}$ convergence probability and the $\overline{C}$ average number of cuts after $N$ iterations steps for the $60\times60$ benchmark max-cut problem also applied in Ref.~\citenum{cai2020}. The red line in panel B shows the maximum number of cuts, i.e. the global solution to the problem. 

It is clear, that at zero noise level, the convergence probability is poor ($\mathbb{P}_\mathrm{conv}=1.5\%$) and the achieved number of cuts is far away from the global solution. As 
the relative amplitude of the dynamic noise is increased, 
the convergence probability (Fig.~\ref{fig:2}A) exhibits a stochastic resonance phenomenon similarly to the results of Ref.~\citenum{cai2020}: irrespective of the noise color $\mathbb{P}_\mathrm{conv}$ shows a peak at $(\Delta G/G)_\mathrm{dynamic} \approx 13.8 \%$ leading to a $\mathbb{P}_\mathrm{conv} = 40-50 \%$ chance of convergence. This figure implies, that at a lower noise level the system sticks to local minima, which prevents the convergence to the global solution, whereas at too high noise level the system is able to escape from the global minimum, which also hampers the convergence. As an interesting conclusion, however, the results of the simulation are very similar for the different noise colors, i.e. the temporal correlations in the noise spectra are irrelevant, and $(\Delta G/G)_\mathrm{dynamic}$ seems a proper, noise-type independent metric to find the optimal noise level. This also allows the simplification of the simulations by easily generated white noise spectra. Furthermore, it is emphasized that the best $(\Delta G/G)_\mathrm{dynamic} \approx 13.8 \%$ noise level corresponds to the top end of the experimentally observed relative noise values (Fig.~\ref{fig:1}A), i.e. the experimentally relevant noise levels do not hamper the network operation, in contrary, experimentally relevant noise levels might not be enough to realize the optimal noise level if stochasticity is solely introduced by the noise of the memristor elements of the crossbar matrix. 

We have repeated these simulations for numerous benchmark problems from the Biq Mac library spanning matrix sizes of $60\times 60$ (circles in Fig.~\ref{fig:2}E), $80\times 80$ (pentagons) and $100\times 100$ (stars), using white, pink and Lorentzian spectra (grey, pink and blue symbols). At the larger matrix sizes, only white noise was applied. For these problems, the symbols in Fig.~\ref{fig:2}E represent the relative noise values, where the convergence probability is maximal. 
Furthermore, we have generated an even larger weight matrix ($300\times 300$, '$+$' symbol in the last column). Here, the global solution is not known, therefore the symbol represents the noise value, where $\overline{C}$ is maximal. Whereas the convergence probability and the $\overline{C}$ strongly vary for the different problems, the optimal noise level scatters around a common $\Delta G/G=13.2\%$ average value (horizontal solid line) with a a small variance of $2.6\%$ (horizontal dashed lines). This analysis does not show any systematic tendencies as a function of the matrix size, even the largest matrix with $90000$ memristor elements exhibits optimal operation close to this average value. We note, that the system size dependence of the optimal noise level was analyzed in Ref.~\citenum{cai2020} as well. However, in the latter analysis the current noise of the entire bit lines was considered. The system-size independent optimal noise level of the individual devices (Fig.~\ref{fig:2}E) yields a bit line current variance scaling with the square root of the array size due to the $\Delta I_j\sim \sqrt{d_j}$ relation (see Eq.~\ref{reloffnoise1}), i.e. the results of Ref.~\citenum{cai2020} on the optimal noise level (Fig.~ 5C in Ref.~\citenum{cai2020}) are consistent with our observations. 

\subsection{Annealing schemes}
\label{annealing}

The greatest challenge with randomization algorithms is that stochasticity helps to escape local minima but there is no guarantee for the system to stay in the global minimum, once it is first reached. A common approach to overcome this difficulty is to "cool" the system, i.e. gradually decrease the extent of stochastic behavior during the optimization process. For an experimentally realized HNN, the straightforward method is to harvest and tune the inherent device noise utilizing the multilevel programmability of the conductance states.

\begin{figure}
    \includegraphics[width= \columnwidth]{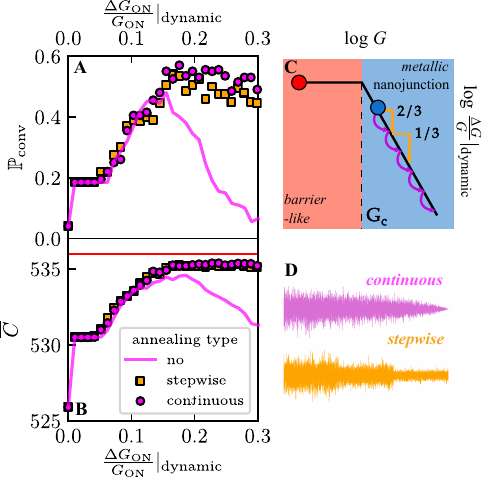}
    \caption{Simulation results for test problem g05\_60.0\,\cite{wiegele2007} using annealed pink dynamic noise. (A) Convergence probability as a function of dynamic noise level for the different schemes: constant noise (pink line), continuous annealing (pink circles), double-step annealing (orange squares). (B) $\overline{C}$ as a function of dynamic noise level for the different schemes (same colors as in (A)). (C) Operation scheme with annealing. A memristor has two operational regimes based on the dynamic noise. The matrix elements representing zero are set to the far OFF state giving essentially zero contribution, whereas matrix elements representing one are programmed to an ON state at the desired initial noise level. During the $N=10^4$ steps at each $K=200$ starting vectors the systems' ON state is gradually reprogrammed to a lower dynamic noise level. (D) Example $G^{(t)}_\mathrm{anneal}$ noise signals for continuous logarithmic and discrete double-step annealing schemes.}
    \label{fig:3}
\end{figure}

According to the work of Cai et. al. \cite{cai2020} the optimal trend for the cooling process in a HNN is superlinear. We have implemented this cooling scheme in our simulations, applying a parameterless superlinear annealing protocol on the stochastic variation of the conductance:
\begin{equation}
        G^{(t)}_\mathrm{anneal} = \log \left( 10 - \frac{9\cdot t}{N} \right) \cdot G^{(t)},
\label{annealeq}    
\end{equation}
where the $G(t)$ noise signal is generated according to the chosen spectrum and the initial $\Delta G/G$ value, and the noise signal is accordingly attenuated as the iterations evolve. The such generated temporal decrease of the noise amplitude is illustrated by the pink curve in Fig.~\ref{fig:3}D.

As illustrated in Fig.~\ref{fig:3}.C: the OFF-state conductance (red dot) is chosen deep in the barrier-like regime (and accordingly $G_\mathrm{OFF}=0$ is applied) while the ON-state (blue dot) is prepared in the metallic regime with non-zero dynamic noise. During the $N$ steps the blue dot is moved towards higher conductances so that the relative dynamic noise gradually decreases (Fig.~\ref{fig:3}.D). All simulations were made using the experimentally motivated pink noise.

The results achieved by this  continuous annealing protocol can be seen in Figs.~\ref{fig:3}A,B scattered as pink circles. Here, the horizontal axis represents the initial noise value. To compare this annealing scheme to the network operation with constant noise, the concerned results from Figs.~\ref{fig:2}A,B are reproduced by pink lines. It is clear, that the annealing procedure started from a high enough noise level delivers significantly better convergence probability than the constant noise simulation using the optimal noise level, which is consistent with the observations in Ref.~\citenum{cai2020}. However, the results plotted in Fig.~\ref{fig:3}A,B demonstrate an unexpected phenomenon: if the annealing is started from a noise level at or below the optimal $13.8\%$ constant noise level, the convergence probability does not show any improvement compared to the related constant noise simulation anymore. This implies an important conclusion: it is not vital to decrease the noise level well below the optimal noise level during the annealing process, however, it is beneficial if the annealing is started from a higher noise level than the optimal constant noise level. In other words, the optimal constant noise level does not cause a significant escape probability from the global solution, however, an initially higher noise level helps to escape from the local minima driving the system more efficiently towards the global solution.

Utilizing this finding, we propose a highly simplified double-step annealing protocol (orange illustrations in panels C and D), where the initial noise level is decreased to its $2/3$ and $1/3$ value at the $1/3$ and $2/3$ of the iterations steps. According to panels A and B, this simplified annealing protocol (orange symbols) delivers similar results as continuous annealing. This is highly beneficial for the network operation, as continuous noise annealing would be a demanding task due to the frequent reprogramming of all memristive cells. The double reprogramming along all the iterations steps is a reasonable trade-off between the time-consuming continuous annealing, and the constant noise operation, where the convergence probability is worse, and it is unrealistic to precisely know the optimal noise level in advance.  

\subsection{Memristive HNN operated in the metallic nanojunction regime}
\label{metallicregime}

\begin{figure}
    \includegraphics[width= \columnwidth]{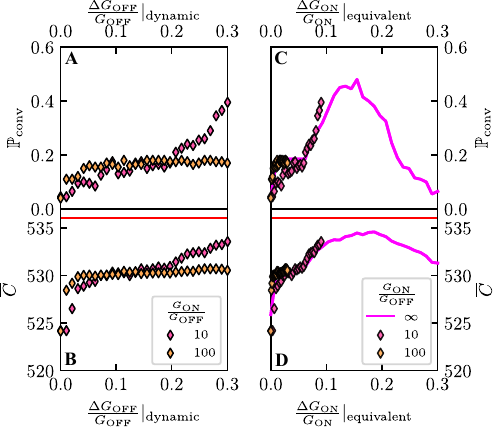}
    \caption{Simulation results for test problem g05\_60.0\,\cite{wiegele2007} operating the noise model in the pure metallic regime, and using pink noise. (A,B) Convergence probability and $\overline{C}$ as a function of the relative dynamic OFF-state noise level (i.e. the dominant noise contribution) using $G_\mathrm{ON}/G_\mathrm{OFF}=10$ and $G_\mathrm{ON}/G_\mathrm{OFF}=100$ conductance ratios (pink and orange). (C,D) The same data rescaled to the equivalent relative ON-state noise level (see text). The pink line reproduces the simulation using solely ON-state noise in the mixed barrier-like and metallic regime (pink data in Figs.~\ref{fig:2}A,B), which actually represents the $G_\mathrm{ON}/G_\mathrm{OFF}=\infty$ limit. Note, the largest $30\%$ relative OFF-state noise levels in panels (A,B) correspond to equivalent noise values of $\approx 0.1$ and $\approx 0.03$ for the $G_\mathrm{ON}/G_\mathrm{OFF}=10$ and $G_\mathrm{ON}/G_\mathrm{OFF}=100$ conductance ratios, respectively.}
    \label{fig:4}
\end{figure}

Along the discussion of Eq.~\ref{reloffnoise3} we have seen that the noise of the OFF-state elements dominates the network once both the OFF and the ON states are positioned in the metallic nanojunction regime, which is described by a $\gamma>1$ exponent. 
Next, we analyze the network operation in this \emph{pure metallic regime} by varying the dominant $\Delta G_\mathrm{OFF}/G_\mathrm{OFF}$ relative noise level using $G_\mathrm{ON}/G_\mathrm{OFF}=10$ (purple symbols in Figs.~\ref{fig:4}A,B) and $G_\mathrm{ON}/G_\mathrm{OFF}=100$ (orange symbols in Figs.~\ref{fig:4}A,B) conductance ratios and $\gamma=3/2$. 
Here, the ON state noise level is also simulated according to the scaling in Eq.~\ref{reloffnoise2}. In this case the convergence probability and $\overline{C}$ (Figs.~\ref{fig:4}A,B) exhibit significantly worse results even at the highest $30\%$ relative noise than the optimal network operation in Fig.~\ref{fig:2}A,B at $13.8\%$ relative ON state noise level. 
This result, however, is obvious from Eqs.~\ref{reloffnoise1}-\ref{reloffnoise2}. According to these formulae, arbitrary OFF and ON state noise levels along the noise model can be converted to an equivalent situation, where the OFF elements are noiseless, but the equivalent relative ON-state noise, $(\Delta G_\mathrm{ON}/G_\mathrm{ON})_\mathrm{equivalent}$ is set such, that the overall current noise of the given bit line remains the same. According to Eq.~\ref{reloffnoise2} the $\gamma=3/2$ and $p_\mathrm{c}=0.5$ parameters yield $(\Delta G_\mathrm{ON}/G_\mathrm{ON})_\mathrm{equivalent}=(\Delta G_\mathrm{OFF}/G_\mathrm{OFF})/ \sqrt{1+G_\mathrm{ON}/G_\mathrm{OFF}}$. In Fig.~\ref{fig:4}C,D the results of Fig.~\ref{fig:4}A,B and Fig.~\ref{fig:2}A,B are plotted as the function of the equivalent ON-state noise level, demonstrating that the curves indeed follow the same tendency. From this we can conclude, that the pure metallic nanojunction regime yields the dominance of the OFF-state elements' noise, however, a given OFF-state noise corresponds to significantly smaller equivalent ON-state noise, i.e. even the largest $30\%$ OFF-state noise is too small to reach the optimal operation. 

\subsection{Further non-idealities}
\label{subsub:finite_off}
\label{furthernonidealities}

After a detailed analysis of the memristive HNNs' noise properties and their impact on the network operation, we analyze the role of further device-nonidealities, like the programming inaccuracy, and the finite $G_\mathrm{OFF}$ conductance. 

\subsubsection{Programming inaccuracy}

\begin{figure}
   \includegraphics[width= \columnwidth]{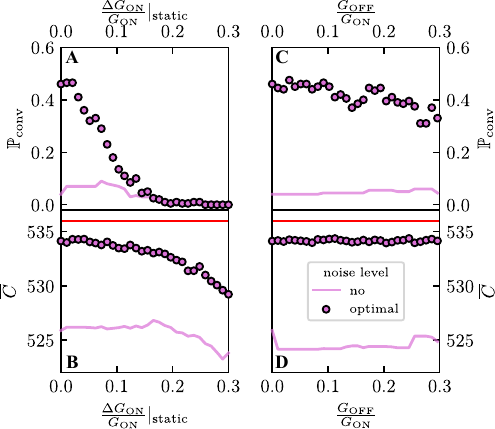}
     \caption{Convergence probability (A,C) and $\overline{C}$ (B,D) as a function of the $(\Delta G/G)_\mathrm{static}$ device-to-device conductance variations (A,B) and the $G_\mathrm{OFF}/G_\mathrm{ON}$ conductance ratio at finite OFF conductance (C,D). 
     Pink circles (lines) represent the results for optimal equivalent dynamic noise level (zero dynamic noise level). Device-to-device variations are modeled by a Gaussian conductance distribution.}
    \label{fig:5}
\end{figure}

In Figs.~\ref{fig:5}A,B we analyze the role of the $(\Delta G/G)_\mathrm{static}$ measure of the programming inaccuracy, i.e. the device-to-device variance of the time-averaged conductance normalized to the average conductance. Here, we also consider the mixed barrier-like and metallic regime using the approximation of $G_\mathrm{OFF}=0$, i.e. solely analyzing the programming inaccuracy of the ON-state conductances. The network's operation for an increasing $(\Delta G/G)_\mathrm{static}$ is demonstrated in Figs.~\ref{fig:5}A,B with no dynamical noise (pink line), and constant pink noise at optimal amplitude (pink circles). 

In the noiseless network device-to-device variations up to $\approx15\%$ leave the poor noiseless network performance practically unchanged, whereas larger $(\Delta G/G)_\mathrm{static}$ already makes the network operation even worse. Here, it is to be emphasized, that static device-to-device variations seemingly produce a stochastic deviation of the bit line currents from the expected values of the original ideal HNN along the temporal evolution of the neural states.\cite{cai2020} However, a finite $(\Delta G/G)_\mathrm{static}$ only deforms the weight matrix of the HNN, but still an ideal noiseless HNN is realized. This means that a finite $(\Delta G/G)_\mathrm{static}$ and $(\Delta G/G)_\mathrm{dynamic}=0$ yields a modified ideal HNN, where the energy can only be reduced along the operation yielding similar dead-ends in the local minima as the original noiseless HNN. Therefore, the device-to-device variations are not proper for performance enhancement in the memristive HNN, for that either true stochasticity (noise) is required, or a non-ideal HNN with somewhat chaotic energy-trajectories should be realized. The latter is possible by the introduction of a diagonal feedback (see Sec.~\ref{diagonalfeedback}), and presumably nonlinear device characteristics also yield similar non-ideal chaotic behavior, the latter, however is not analyzed in this paper.

It is also interesting to analyze the role of the programming inaccuracy if it is accompanied by optimal dynamical noise characteristics. According to the pink circles in Figs.~\ref{fig:5} already $(\Delta G/G)_\mathrm{static}>0.025$ values yield a sharp decrease in the convergence probability. An annealed network would show a very similar decay of the convergence probability (not shown). Accordingly, the proper programming accuracy is vital in the network. Such accuracy has already been experimentally demonstrated in memristors with 2048 distinct conductance levels (corresponding to 11-bit resolution), where a special denoising process was applied to maximize programming accuracy.\cite{Rao2023} The states were programmed between 4144~$\mu$S and 50~$\mu$S, with a 2~$\mu$S resolution, which roughly estimates to $\Delta G / G\approx 0.0005-0.04$, i.e. if the network is operated in the high conductance end of this conductance regime, the envisioned $(\Delta G/G)_\mathrm{static}<0.025$ condition (see Fig.~\ref{fig:5}A) is easily satisfied even at much worse conductance resolution. 

\subsubsection{Finite OFF conductance}
\label{finiteoff}
We have seen that from the noise perspective one can always find an equivalent picture, where the OFF state is noiseless, i.e. in this sense the partitioning of the noise between the ON and OFF elements is irrelevant, just the overall noise matters. However, even at zero noise, a finite OFF-state conductance may modify the network operation due to the imperfect representation of zero states.
To analyze this, simulations were run at different $G_\mathrm{OFF}/G_\mathrm{ON}$ values - ranging from $0$ to $0.3$ - with no dynamical noise, and constant pink noise with optimal equivalent amplitude. The results can be seen in Figs.\ref{fig:5}C,D. No significant change is detected in a noiseless network (pink line), but in a network with optimal equivalent noise the finite, $(G_\mathrm{OFF}/G_\mathrm{ON})>0.1$  values already yield shallow but significant reduction of the convergence probability (pink circles).

\subsection{Externally induced stochasticity in memristive HNNs with suboptimal internal noise level}
\label{external}

\subsubsection{External injection of current noise}

The above considerations have demonstrated, that rather large, $\approx 11-16\%$ relative equivalent ON-state noise levels are required for the best network operation, which can be further boosted by annealing the noise from an even higher initial level. These noise levels are already at the border of the experimentally observed noise values, and especially in the pure metallic regime it is hardly possible to reach the optimal noise level in the network. On the other hand this also means that the network is easily set to an operation regime, where the overall noise is definitely smaller than the optimal level, i.e. in this regime it is possible to apply external noise injection with which the stochastic operation is optimized. This scheme is demonstrated in Figs.~\ref{fig:6}A,B (see also Fig.~\ref{fig:0}A,D), where the light green arrows illustrate noise injection to the bit-line current from an external tunable noise source.  Mathematically this is represented by the proper modification of the update rule (Eq.~\ref{activation}):
\begin{equation}
       x_j^{(t+1)} =
\begin{cases}
+1 \quad \mbox{if} \ a_j^{(t)} \geq \theta_j+\xi_j^{(t)} \text{,} \\
-1 \quad \mbox{if} \ a_j^{(t)} < \theta_j+\xi_j^{(t)} \text{,}
\end{cases}
\label{activation2}
\end{equation}
where $\xi_j^{(t)}$ is a stochastic variable with $\sigma_j$ standard deviation representing the external noise injection (note, that for the max-cut problem $\theta_j=0$ applies).  As proposed in Ref.~\citenum{cai2019harnessing}, an additional memristive crossbar line with high amplitude tunable noise characteristics could be applied to tailor the noise level in the bit lines separately. Here, we apply an even simpler scheme with the single external memristive (or non-memristive) tunable noise source representing a $\sigma$ standard deviation of the $\xi$ random variables. Along the multiplexing this external noise is added to the randomly chosen bit line. The green line (green symbols) in Figs.~\ref{fig:6}A,B represent the convergence probability and $\overline{C}$ as a function of $\sigma$ for constant (annealed) external noise amplitudes. These curves highly resemble the results, where a constant (annealed) internal noise of the crossbar elements was applied (Figs.~\ref{fig:2}A,B and \ref{fig:3}A,B). This is, however, an easily deducible correspondance, as the $\sigma_j$ standard deviation of the external noise can be converted to an $(\Delta G_\mathrm{ON}/G_\mathrm{ON})_{\mathrm{equivalent},j}=\sigma_j/\sqrt{d_j}$ ON-state equivalent noise level in bit line $j$ considering the mixed barrier-like and metallic regime with neglected OFF-state noise. Using this conversion factor with $\sigma_j=\sigma$ and replacing $d_j$ by its average value, we can plot the results of Figs.~\ref{fig:2}A,B and \ref{fig:3}A,B on Fig.~\ref{fig:6} (see the rescaled top horizontal axis). With this rescaling the constant and annealed internal memristor noise (purple line and purple circles in Fig.~\ref{fig:6}) indeed exhibits highly similar effect on the network performace, as the equivalent externally injected constant or annealed noise (green line and green circles). 

\subsubsection{Stochastic variation through diagonal feedback}
\label{diagonalfeedback}

\begin{figure}
\includegraphics[width= \columnwidth]{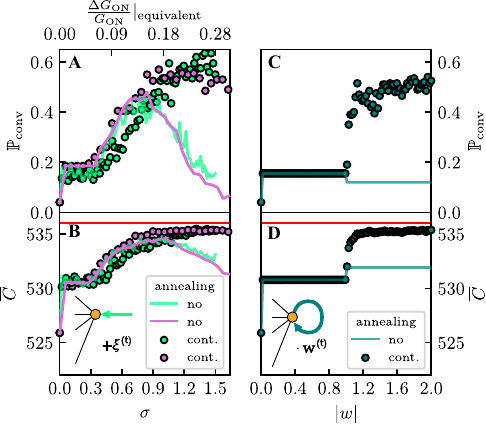}
\caption{(A,B) The effect of external noise injection on the performance of the memristive HNN using the simulation results for the  g05\_60.0\,\cite{wiegele2007} benchmark problem. The green line (circles) represent the convergence probability (A) and $\overline{C}$ as the function of the external noise's standard deviation, $\sigma$. As a reference, $\mathbb{P}_\mathrm{conv}$ and $\overline{C}$ are also plotted for the case of constant and annealed internal noise (purple line and circles) using the rescaled, equivalent ON-state noise axis (see top axes). For the external noise injection white noise was applied, and the annealing protocol (circles) followed the continuous annealing scheme (Eq.~\ref{annealeq}). (C,D) The effect of diagonal self-feedback on the performance of the same memristive HNN using finite negative $w$ values and zero noise. $\mathbb{P}_\mathrm{conv}$ and $\overline{C}$ are plotted as a function of $|w|$ for constant $w$ (blue lines) and for continuously annealed $w$ (see Eq.~\ref{annealeq}). The insets in (A) and (C) illustrate the external noise injection and diagonal feedback schemes similarly to Fig.~\ref{fig:0}A.}
   \label{fig:6}
\end{figure}

In Ref.~\citenum{cai2020} tunable stochasticity was introduced by another strategy, which relied on the internal noise of the network, which was attenuated or magnified by a tunable hysteretic threshold circuitry. The latter practically induced a self-feedback to the network (see the dark green arrow in Fig.~\ref{fig:6}C and also in Fig.~\ref{fig:0}A), which is equivalent to the presence of non-zero elements at the diagonal of the weight matrix. Mathematically this modifies the update rule as:
\begin{equation}
       x_j^{(t+1)} =
\begin{cases}
+1 \quad \mbox{if} \ a_j^{(t)} + w\cdot x_j^{(t)} \geq \theta_j \text{,} \\
-1 \quad \mbox{if} \ a_j^{(t)} + w\cdot x_j^{(t)} < \theta_j \text{,}
\end{cases}
\label{selffeedback}
\end{equation}
where $a_j^{(t)}$ is the activation without self-feedback, and $w$ is the tunable strength of the self-feedback. Positive $w$ values drive the network towards the stabilization of the actual states, i.e. if. a certain neuron would change its state at $w=0$, it is possible that at $w>0$ it does not change the state (for $+1$ ($-1$) neural state the state-change would require negative (positive) activation, but the self-feedback shifts the threshold towards positive (negative) values). With similar argumentation, a negative $w$ yields neural updates in situations, where the ideal Hopfield network ($w=0$) would not yield the change of the neural state. It can be shown, that the latter introduces a chaotic behavior to the network, which can be utilized to escape from local minima.\cite{yang2020} In a memristive crossbar array, the introduction of negative weights would require the application of differential memristor pairs,\cite{Li2018b} i.e. it is more reasonable to keep the memristive matrix positive valued with zero diagonal elements, and to introduce the self-feedback as an offset along the neural updates. 

Here we wish to investigate the effect of the self-feedback separately from the effect of internal noise, i.e. we have performed simulations for a noiseless Hopfield network using either constant (blue lines in Figs.~\ref{fig:6}C,D), or annealed (blue circles in Figs.~\ref{fig:6}C,D) negative $w$ values. The latter exhibits similar convergence probabilities as the case of annealed internal or externally injected noise (purple and green circles Figs.~\ref{fig:6}A), however, at constant self-feedback amplitude (blue line in Fig.~\ref{fig:6}C) no optimal $w$ is found, which would yield similarly good convergence probability ($\approx 45\%$), as the optimal constant internal or external noise level (purple and green lines in Fig.~\ref{fig:6}A), rather the convergence probability remains below $\approx 20\%$ regardless of the $w$ value. Accordingly, we can state the the annealing is vital in the case of self-feedback.

\section{Conclusions}

In conclusion, we simulated probabilistic optimization schemes inspired by the memristive HNNs experimentally realized in Ref.~\citenum{cai2020,cai2019harnessing}. These works demonstrated, that memristive HNNs are not only efficient hardware accelerators for complex optimization tasks thank to their single-step matrix-vector multiplication capability, but the intrinsic noise of the memristive elements can be exploited as a hardware resource, introducing proper stochasticity to the network. As a main focus we simulated the operation of the memristive HNNs relying on experimentally deduced, realistic noise characteristics. Based on a broad range of conductance dependent noise characteristics in various memristive systems, we proposed a noise model describing the typical noise evolution along the variation of the conductance states. Relying on this model, we demonstrated distinct operation regimes, where either the ON-state or the OFF-state noise provides dominant contribution. We also demonstrated, that the relative conductance variation is not only a good measure of the noise amplitude, but is a highly relevant parameter describing the operation of the memristive HNNs: according to our simulations the relative noise level required for the optimal network operation is found to be in the range of $\Delta G/G\approx 11-16\%$ regardless of the color of the noise spectrum (white, pink or Lorentzian noise) or the size of the problem ($60\times 60 - 300 \times 300$). We have shown, that further performance enhancement can be achieved by noise annealing, however, a highly simplified and easily implemented double step noise annealing scheme provides similar performance, as the more refined, continuous super-linear noise annealing scheme. It is also found, that the optimal noise level is at the top edge of the experimentally achievable relative noise levels, which means, that the network is easily tuned to an operation regime with suboptimal relative noise level, where the optimal operation can be either set by external noise injection or by a negative diagonal feedback, introducing a chaotic network behavior. Finally, we have explored the effects of further non-idealities, such as the limited programming accuracy, and the finite OFF-state conductance of the memristors.  We have argued that any static non-ideality that deforms the weight matrix but still implements a noise-free HNN can only lead to a degradation of the network performance, i.e. for performance enhancement either true stochasticity (noise) or a non-ideal HNN with somewhat chaotic energy trajectories is required. It was, however, also found, that static non-idealities, especially the device-to-device variations with $(\Delta G/G)_\mathrm{static}>0.025$ cause severe performance degradation in networks with optimized dynamical noise level. 

The rapidly growing field of memristor research is expected to deliver radically new IT solutions in the nearest future. We believe that our results contribute to this field, by exploring the prospects of fully connected memristive networks utilizing the inherent stochasticity of memristors for probabilistic optimization algorithms.

\begin{acknowledgments}
This research was supported by the Ministry of Culture and Innovation and the National Research, Development and Innovation Office within the Quantum Information National Laboratory of Hungary  (Grant No. 2022-2.1.1-NL-2022-00004), the \textbf{ÚNKP-22-2-I-BME-73} and \textbf{ÚNKP-22-5-BME-288} New National Excellence Program of the Ministry for Culture and Innovation from the source of the National Research, Development and Innovation Fund and the NKFI K143169 and K143282 grants. Project no. 963575 has been implemented with the support provided by the Ministry of Culture and Innovation of Hungary from the National Research, Development and Innovation Fund, financed under the KDP-2020 funding scheme.
Z.B. acknowledges the support of the Bolyai János Research Scholarship of the Hungarian Academy of Sciences.
The authors are grateful to Dávid Krisztián and Péter Balázs for their contribution to the noise measurements on SiO\textsubscript{x} resistive switches.
\end{acknowledgments}

\section*{Author Declarations}

\subsection*{Conflict of Interest}

The authors have no conflicts to disclose.

\subsection*{Author Contributions}

The program codes were developed by J.G.F. with supporting contribution from T.N.T. The simulations were run and the data analysis was performed by J.G.F and the work was revised by Z.B.
The noise measurements on SiO\textsubscript{x} were performed by former graduate students and J.G.F under the supervision of Z.B and A.H.
Model calculations were performed by J.G.F, Z.B. and A.H. The manuscript was written by J.G.F. and A.H. All authors contributed to the discussions. The project was supervised by A.H. with support from B.Z. and T.N.T.

\section*{Data Availability Statement}

The data that support the findings of this study are available from the corresponding author upon reasonable request.

\bibliography{aipsamp}

\end{document}